\documentclass[sigconf, authorversion, nonacm, 9pt]{acmart}

\usepackage{fancyhdr}
\AtBeginDocument{%
	
	\fancyhead[C]{\fontsize{8}{8} \selectfont CC-BY 4.0. This is the author’s preprint version of the camera-ready article. A full version of this paper is published in the Proceedings of the 22nd International Middleware Conference: Demos and Posters (Middleware '21).}%
	
	\addtolength{\footskip}{2.0\baselineskip}%
	
	\fancyfoot[C]{\fontsize{8}{8} \selectfont Original ACM publication available at: https://dl.acm.org/doi/abs/10.1145/3491086.3492472.}%
}

\renewcommand\footnotetextcopyrightpermission[1]{} 
\AtBeginDocument{%
  \providecommand\BibTeX{{%
    \normalfont B\kern-0.5em{\scshape i\kern-0.25em b}\kern-0.8em\TeX}}}

\begin{document}

\title[]{A Milestone for FaaS Pipelines, Object Storage- vs VM-Driven Data Exchange}

\author{Germán T. Eizaguirre}
\email{germantelmo.eizaguirre@urv.cat}
\affiliation{%
	\institution{Universitat Rovira i Virgili}
	\city{Tarragona}
	\country{Spain}
}

\author{Marc Sánchez-Artigas}
\email{marc.sanchez@urv.cat}
\affiliation{%
	\institution{Universitat Rovira i Virgili}
	\city{Tarragona}
	\country{Spain}}

\author{Pedro García-López}
\email{pedro.garcia@urv.cat}
\affiliation{%
	\institution{Universitat Rovira i Virgili}
	\city{Tarragona}
	\country{Spain}
}

\renewcommand{\shortauthors}{Eizaguirre, et al.}

\begin{abstract}
Serverless functions provide high levels of parallelism, short start-up times, and ``pay-as-you-go'' billing. These attributes make them a natural substrate for data analytics workflows. However, the 
impossibility of direct communication between functions makes the execution of workflows challenging. The current practice to share intermediate data among functions is
through remote object storage (e.g., IBM COS). Contrary to conventional wisdom, the performance of object storage is not well understood. For instance, object storage can even be superior to other
simpler approaches like the execution~of shuffle stages (e.g., \texttt{GroupBy}) inside powerful VMs to avoid all-to-all transfers between functions.                                                                                                       
Leveraging a genomics pipeline, we show that object storage is a reasonable choice for data passing when the appropriate number of functions is used in shuffling stages.
\end{abstract}


\let\oldmaketitle\maketitle
\renewcommand{\maketitle}{%
	\oldmaketitle%
	\thispagestyle{fancy}}
\maketitle

\section{Problem Statement and Goals.}

Serverless computing platforms (such as AWS Lambda, IBM Cloud Functions, etc) provide on-demand scalability and fine-grained resource allocation. 
This means that a developer~can leverage massive levels of parallelism in just a few seconds to build applications with rapid horizontal scaling.
Along with their ease-of-use, these platforms have been recently used to execute serverless workflows composed of a sequence of execution stages, which
can be represented as a directed acyclic graph (DAG). DAG nodes correspond to serverless functions and edges correspond to  the flow of data between dependent
stages.

Unfortunately, serverless functions do not support point-to-point communication. 
As a result, the standard practice for passing intermediate data between serverless functions~is 
through remote object storage (e.g., IBM COS). If care is not taken, however, I/O-bound stages that require all-to-all data transfers
between functions (such as \texttt{GroupBy} and \texttt{OrderBy}) can end up bottlenecking the system. This typically occurs due to
the limited throughput of object storage services (e.g., IBM COS only supports a few thousand operations/s). 

On the positive side, object storage is cheap and an ``always-on'' service, requiring little intervention from the user. It is therefore the most comfortable option 
for programmers and data analysts, despite its higher latency and lower throughput compared to other alternatives such as AWS ElastiCache. 
For this reason, some practitioners prefer to run I/O-bound stages inside large-memory virtual machine (VM) instances to minimize data transfers. Nevertheless, these solutions 
do not exploit the huge aggregated bandwidth offered by object stores.

The objective of this demo is to show the practical utility of a pure serverless implementation for workflows. That is,
we demonstrate that object storage performs well when the appropriate number of functions is used in I/O-bound stages.
To do so, we run a genomics pipeline in two manners. One way that is ``purely'' serverless using object storage,
and the other when the shuffle operation runs inside a powerful VM. In this way, we qualitatively evaluate the pros and cons of each strategy
for serverless workflows. 

\section{Research and Technical Approach.}

\subsection{Pipeline Description.}

We employ METHCOMP~\cite{10.1093/bioinformatics/bty143}, a compression method for DNA methylation annotation files, as an evaluation workflow, since it is habitual
in the genomics community. For instance, 
the ENCODE project repository contains bisulfite data for more than $500$ samples. Unfortunately, raw data in structured BED format\footnote{\href{https://www.encodeproject.org/data-standards/wgbs/}{https://www.encodeproject.org/data-standards/wgbs/}} can amount to tens of GBs. METHCOMP presents a compression method tailored to methylation data that yields about $10$x better compression ratio than \texttt{gzip}. Particularly, this method operates in two consecutive stages. A first \texttt{sort} stage that entails
all-to-all data transfers between functions, and a second stage that is embarrassingly parallel (encoding). 
For the purpose of this demonstration, we port the METHCOMP pipeline to serverless.

\subsection{Pipeline Implementations.}

We use Lithops~\cite{10.1109/MS.2020.3029994}, an open source\footnote{\href{https://github.com/lithops-cloud/lithops}{https://github.com/lithops-cloud/lithops}} Python framework for serverless analytics. Lithops allows the parallel execution of analytics workflows on top of cloud functions. Moreover, it supports the provisioning of heavyweight VM instances to run computations, which gives us the chance  to seamlessly run different incarnations of the same pipeline. We use IBM Cloud as the cloud provider for both scenarios. IBM COS is used as data passing mechanism in both pipelines.

\begin{figure}[h]
	\centering
	\includegraphics[width=0.9\linewidth]{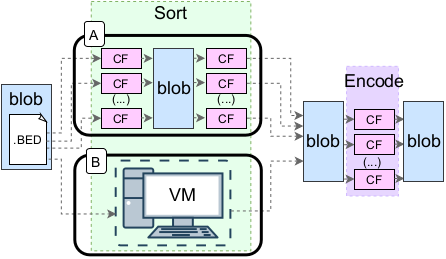}
	\caption{Purely serverless (A) and hybrid implementations (B) of the genomics compression pipeline.}
	\Description{Purely serverless (upper branch) and hybrid implementations (lower branch) of the genomics compression pipeline.}
	\label{fig:confs}
\end{figure}

Our first implementation is VM-based (Figure~\ref{fig:confs}, A). Thanks to Lithops VM provisioning, we execute the \texttt{sort} stage within a VM with sufficient physical memory, 
while resorting to cloud functions only for the encoding stage.

Our second implementation is purely serverless (Figure~\ref{fig:confs}, B). For the \texttt{sort} operation, we use Primula~\cite{10.1145/3429357.3430522}, an extension of Lithops that optimizes 
all-to-all transfers between functions. Along with a number of I/O optimizations for serverless all-to-all communication, Primula finds the optimal number of functions for a given shuffle data size ``on the fly''. 
For I/O-bound tasks, using the optimal number of functions in terms of remote storage resource utilization is crucial for good performance~\cite{10.1145/3429357.3430522}.

\subsection{Experiment setup and results}

In the demo, we will evaluate both configurations in the \texttt{us-east} region using the $3.5$GB dataset: ENCFF988BSW\footnote{\href{https://www.encodeproject.org/experiments/ENCSR515MHO/}{https://www.encodeproject.org/experiments/ENCSR515MHO/}}. We will allocate $2$GB of memory to cloud functions, and use a bx2-8x32 IBM Virtual Server Instance as our VM for the \texttt{sort} stage.

Table ~\ref{tab:res} shows our results in terms of end-to-end latency and cost for a parallelism degree of $8$ workers.
End-to-end latency includes startup times. Cost subsumes the following charges: the cost of cloud functions, storage requests, 
and the VM expenses --- i.e., execution time and storage volume. While both configurations deliver similar costs, the ``purely'' serverless architecture 
significantly outperforms the hybrid pipeline in terms of latency. This shows that the execution of DAGs over a serverless infrastructure 
can be superior to a VM-based, ``serverful'' infrastructure thanks to a better exploitation of the aggregated bandwidth offered by object storage
when sharing intermediate data.

\begin{table}
	\caption{Performance of METHCOMP pipeline in terms of latency and cost for a $3.5$GB input.}
	\label{tab:res}
	\begin{tabular}{lrr}
		\toprule
		Configuration &Latency (s)&Cost (\$)\\
		\midrule
		"Purely" serverless & 83.32 & 0.008\\
		VM-supported & 142.77 & 0.010\\
		\bottomrule
	\end{tabular}
\end{table}

\subsection{User Interface and Configuration.}
\label{usability}

As our main objective is to show the feasibility of executing workflows using a pure serverless
architecture, we found it imperative to provide a declarative programming interface to represent workflows. 
To this end, we augmented Lithops with a module to create pipelines from JSON configuration files.
Further, we also developed a IPython interface for job tracking in real time, which displays the workflow progress 
and breaks the cost down at each stage.

\section{Related work}

Research on serverless analytics systems is gaining attraction in the last few years. Lithops~\cite{10.1145/3284028.3284029} is an open-source, general-purpose tool for serverless analytics. 
Lithops is now multi-cloud~\cite{10.1109/MS.2020.3029994} and for legacy support, it also enables the execution of Python code as is inside VM instances. To improve the performance of
all-to-all data transfers through object storage, Lithops integrates Primula~\cite{10.1145/3429357.3430522}.

\begin{acks}
This work has been partially supported by EU Horizon $2020$ (No. $825184$) and by the 
Spanish Government (PID2019-106774RB-C22).  Marc S\'{a}nchez-Artigas is a Serra H\'{u}nter Fellow.
\end{acks}

\bibliographystyle{ACM-Reference-Format}
\bibliography{sample-base}

\end{document}